\documentclass[prl,twocolumn,amsmath,showpacs]{revtex4-1}
\usepackage{graphicx,amssymb}
\usepackage[usenames,dvipsnames]{xcolor}
\usepackage[FIGTOPCAP]{subfigure}

\begin{document}

\title{Influence of impurity spin dynamics on quantum transport in epitaxial graphene}

\author{Samuel Lara-Avila}
\affiliation{Department of Microtechnology and Nanoscience, Chalmers University of Technology, G\"{o}teborg, S-412 96, Sweden}

\author{Oleksiy Kashuba}
\affiliation{Institute of Theoretical Physics, Technische Universit\"at Dresden, Dresden, 01062, Germany}

\author{Joshua A. Folk}
\affiliation{Quantum Matter Institute, University of British Columbia, Vancouver BC, V6T 1Z4, Canada}
\affiliation{Department of Physics and Astronomy, University of British Columbia, Vancouver BC, V6T 1Z1, Canada}

\author{Silvia L\"{u}scher}
\affiliation{Quantum Matter Institute, University of British Columbia, Vancouver BC, V6T 1Z4, Canada}
\affiliation{Department of Physics and Astronomy, University of British Columbia, Vancouver BC, V6T 1Z1, Canada}

\author{Sergey Kubatkin}
\affiliation{Department of Microtechnology and Nanoscience, Chalmers University of Technology, G\"{o}teborg, S-412 96, Sweden}

\author{Rositza Yakimova}
\affiliation{Department of Physics, Chemistry and Biology (IFM), Link\"{o}ping University, Link\"{o}ping, S-581 83, Sweden}

\author{T.J.B.M. Janssen}
\affiliation{National Physical Laboratory, Teddington, TW11 0LW, UK}

\author{Alexander Tzalenchuk}
\email[]{alexander.tzalenchuk@npl.co.uk}
\affiliation{National Physical Laboratory, Teddington, TW11 0LW, UK}
\affiliation{Royal Holloway, University of London, Egham, TW20 0EX, UK}

\author{Vladimir Fal'ko}
\affiliation{Physics Department, Lancaster University, Lancaster, LA1 4YB, UK}

\date{version 16}

\begin{abstract}
Experimental evidence from both spin-valve and quantum transport measurements points towards unexpectedly fast spin relaxation in graphene. We report magnetotransport studies of epitaxial graphene on SiC in a vector magnetic field showing that spin relaxation, detected using weak-localisation analysis, is suppressed by an in-plane magnetic field, $B_{\parallel}$, and thereby proving that it is caused at least in part by spinful scatterers.  A non-monotonic dependence of effective decoherence rate on $B_{\parallel}$ reveals the intricate role of scatterers' spin dynamics in forming the interference correction to conductivity, an effect that has gone unnoticed in earlier weak localisation studies.
\end{abstract}

\maketitle

Loss of quantum information carried in the phase and spin of electrons propagating in a disordered conductor is associated with decoherence that suppresses interference corrections to conductivity \cite{AKLL,Larkin}. The fundamental relation between spin relaxation and the low-temperature magnetoresistance (MR) has been investigated theoretically and experimentally in numerous disordered metallic and semiconductor structures.

Recent studies of spin-valve \cite{Yang,Maassen,Dlubak,Maassen2013} and quantum transport \cite{Lara,Jobst,Kozikov,Folk,Haug} effects in graphene-based devices have returned unexpectedly fast spin relaxation in graphene, despite this one-carbon-atom-thin material being billed as an ideal medium for spintronics applications \cite{Avsar,Pesin,Han}. Spin relaxation can be induced by either spin-orbit coupling or spin-flip scattering due to magnetic impurities. In quantum transport, spin-orbit coupling inverts the conventional negative MR around zero perpendicular field, $B_\perp$, (weak localization, WL) to positive MR known as weak anti-localisation, but this effect has never been observed in pristine exfoliated or epitaxial graphenes \cite{Savchenko2008,Savchenko2009,Lara,Jobst,Kozikov,Folk}. By contrast, spin-flip scattering on magnetic impurities washes out quantum interference effects \cite{Lee,Stone,Chandrasekhar,Geim,Levy}, including weak (anti-)localization.  Spin-flip scattering in disordered metals has been shown \cite{Birge,Birge1,Birge2} to raise the electron decoherence rate  $\tau_\varphi^{-1}(T)$ above the value expected from inelastic scattering on thermal charge fluctuations \cite{BLA},
\begin{equation} \label{eq:AA}
\tau_T^{-1} = (k_B T/\hbar)(\rho e^2/h) \ln (h/2e^2\rho) \equiv AT,
\end{equation}
where $\rho$ is resistivity.   Residual decoherence in the limit $T\rightarrow 0$ is then determined only by the electron spin relaxation time, $\tau_s$.  Spin-flip decoherence was previously observed in exfoliated graphene \cite{Folk}. Applying a large in-plane magnetic field, $B_\parallel\gtrsim B_T\equiv k_B T/g_i\mu_B$, polarises impurity spins with g-factor $g_i$, suppressing spin-flip scattering and prolonging phase coherence \cite{Folk}.   For magnetic fields too small to polarize impurity spins, $B_{\parallel} < B_T$, one might expect minimal effect on decoherence.

In this paper we show both experimentally and theoretically that such small in-plane fields do in fact have a discernable effect on phase coherence, but the effect is opposite to that observed for larger $B_\parallel$.  The measurement is performed on epitaxial graphene grown on silicon carbide (SiC/G), using  curvature of the $B_\perp$ MR peak to quantify the electron decoherence rate.
Applying an in-plane magnetic field first broadens the MR peak slightly (enhances decoherence), before the sharpening effect due to impurity polarization sets in. This magnetic field dependence shows that the observed decoherence is caused by spin-flip scattering rather than other dynamical sources of decoherence, such as external noise due to external two-level systems \cite{othernoise}. The non-monotonic dependence of decoherence rate on $B_\parallel$ has not, to our knowledge, been discussed in previous work. It is a generic feature of quantum transport in disordered conductors that can be attributed to the precession of impurity spins at the frequency difefrent from the spin precession of mobile electrons.

When electron ($e$) and impurity ($i$) $g$-factors differ, the difference between their spin precession frequencies,  $\omega_{e\vert i}=g_{e\vert i}\mu_B B_{\parallel}/\hbar$, leads to a random variation of the spin-dependent scattering conditions for electron waves retracing the same closed diffusive trajectory in clockwise and anti-clockwise directions, whose interference forms the quantum correction to conductivity.  The non-monotonicity is characterized by a magnetic field scale, $B_* = \hbar\tau_s ^{-1}/|g_e-g_i|\mu_B$,  above which the decohering effect of the in-plane field (separating precession frequencies for impurities and conduction electrons) is overcome by polarization of impurity spins.  The $g$-factor of magnetic scatterers  can thus be determined by fitting the temperature and  $B_{\parallel}$ dependence of MR curvature to theory developed below.

\begin{figure}[b]
\begin{center}

\subfigure[]{
\includegraphics [scale=0.7] {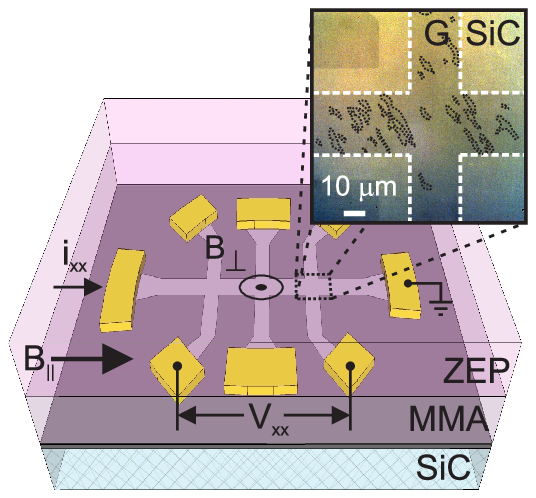}
\label{fig:subfig1a}
}\\
\subfigure[]{
\includegraphics [scale=0.65] {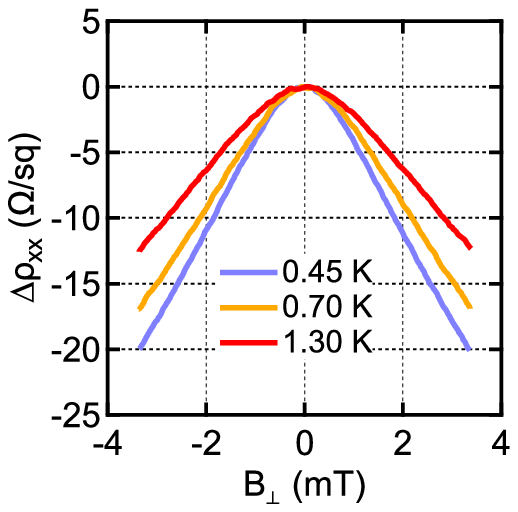}
\label{fig:subfig1b}
}
\subfigure[]{
\includegraphics [scale=0.65] {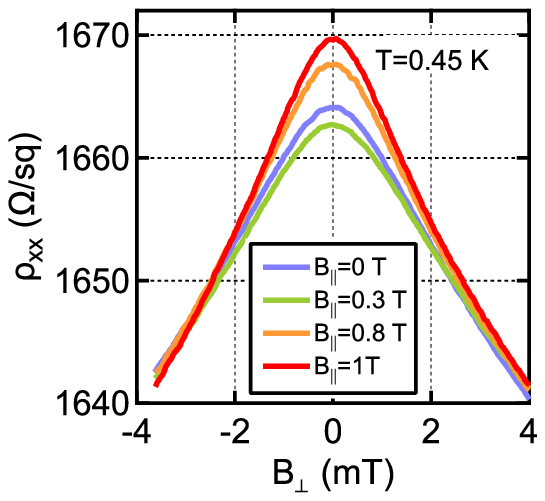}
\label{fig:subfig1c}
}

\end{center}
\caption{(a) Layout of the sample and an optical micrograph \cite{Yager} showing predominantly monolayer graphene with bilayer inclusions; (b) Magnetoresistance sharpens with decreasing temperature ($B_\parallel =0$); (c) The sharpness of the magnetoresistance curve is non-monotonic in $B_\parallel$ ($T=450$ mK).} \label{fig:1}
\end{figure}

Magnetotransport measurements were performed on a large area ($160~\mu$m $\times~35~\mu$m, Fig. \ref{fig:1}(a)) n-doped SiC/G Hall bar, encapsulated in a polymer to improve temporal stability and doping homogeneity and exposed to deep-UV light  to reduce carrier concentration ($n=10\pm1\times 10^{11}~\rm{cm}^{-2}$) \cite{LaraAvila2010}. The device was measured in a dilution refrigerator equipped with a two-axis magnet, allowing independent control of $B_{\perp}$ and $B_{\parallel} $\cite{Folk}.  Average MR measurements are not obscured by mesoscopic conductance fluctuations due to the large sample size.

Figure \ref{fig:1}(b) shows the characteristic negative MR of weak localisation, measured for $B_\parallel=0$.  As expected, MR is sharpest at the lowest temperatures where thermal charge fluctuations are minimized [Eq.~(\ref{eq:AA})].    Raising the in-plane field to $B_\parallel=1$~T yields a significantly sharper (and higher) MR peak due to the partial polarization of impurity spins at this field [Fig.~\ref{fig:1}(c)].  Even in the raw data of Fig.~\ref{fig:1}(c), however, the non-monotonicity that is the primary subject of this paper can be clearly seen: the MR peak for $B_\parallel=0.3$~T is broader than the trace at $B_\parallel=0$, despite the small but non-zero impurity polarization at this low in-plane field.

\begin{figure}[t]

\begin{center}

\subfigure[]{
\includegraphics [scale=0.65] {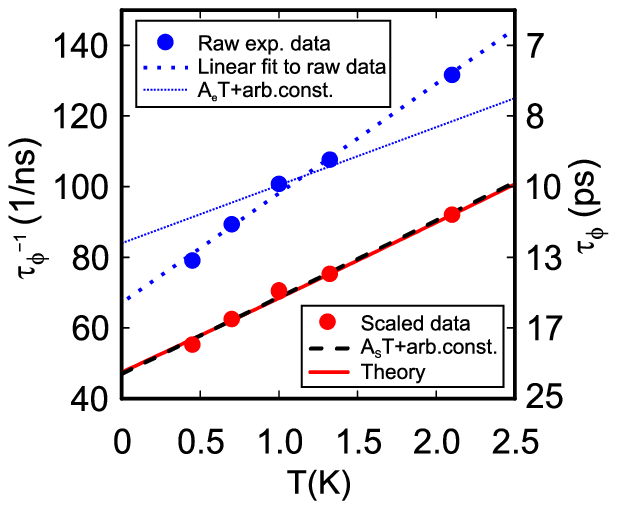}
\label{fig:subfig2a}
}
\subfigure[]{
\includegraphics [scale=0.65] {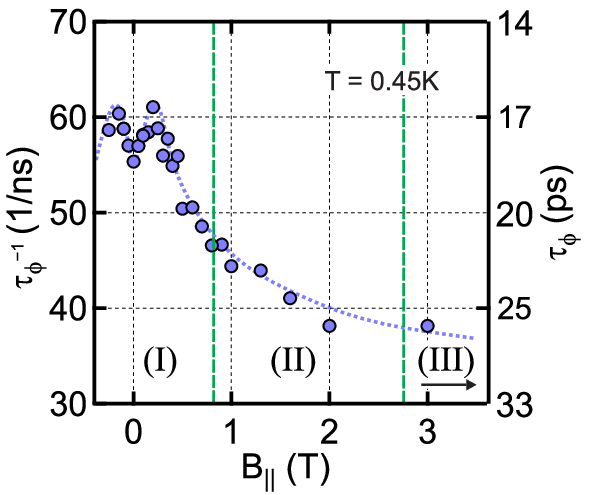}
\label{fig:subfig2b}
}

\end{center}
\caption{(a) Dephasing rate as a function of temperature at $B_\parallel =0$ extracted from the raw (blue circles) and rescaled (red circles) magnetoresistance. The blue dotted line is the best linear fit to the raw data. The slopes of the dashed lines are given by Eq. (\ref{eq:AA}), and the offset constants are chosen arbitrarily for ease of comparison with the data. The solid red line is the theory developed in this work. (b) Dephasing rate extracted from the rescaled magnetoresistance plotted as a function of $B_\parallel$ at $T=450$~mK. Roman numerals denote different scattering regimes. The dotted line is a guide to the eye.} \label{fig:2}
\end{figure}

To make further progress, the decoherence time, $\tau_{\varphi}$, is quantified using the expression for the curvature $\kappa$ of magneto-conductivity around $B_{\perp}=0$,
\begin{equation} \label{eq:1}
\left. \kappa \equiv \frac{\partial ^2 \sigma}{\partial B_{\perp}^2} \right|_{B_{\perp}=0} = \frac{16\pi}{3}  \frac{e^2}{h}
\left( \frac{D\tau_{\varphi}}{h/e} \right)^2
\end{equation}
that comes from the basic functional form of WL \cite{AKLL,Larkin,Note1}. Curvature is extracted from a parabolic fit to the average of 10 measurements of $\rho(B_\perp)$ covering the range of $| B_\perp | \leq 0.5$ mT. The resulting temperature dependence of $\tau_\varphi^{-1}(T)$ at $B_\parallel=0$ clearly shows the linear scaling expected from Eq.~(\ref{eq:AA}) [Fig.~\ref{fig:2}(a)].

The slope $A$ of the temperature dependence in Eq.~(\ref{eq:AA}), estimated using $\rho_{xx}=1500\pm35~\Omega$ which we determine from the measured resistance and the sample aspect ratio yields $A_e=16.4~\rm{K}^{-1}\rm{ns}^{-1}$ as compared to $A \approx 31~\rm{K}^{-1}\rm{ns}^{-1}$ fitted to experimentally-determined $\tau_\varphi^{-1}(T)$ [Fig.~\ref{fig:2}(a)]. This difference can be reconciled in two ways:  (i) To note that the large-area epitaxial graphene monolayers usually contain bilayer inclusions, which we also identified in the device used here, [Fig.~\ref{fig:1}(a)], and which have much higher conductivity than that of the monolayer material \cite{Giannazzo,Kazakova,Carbon}, hence, reducing  the effective length of the Hall bar.  (ii) To treat parameter $A$ as an empirical factor \cite{Savchenko2008}. For the purposes of this paper we simply rescale the effective sample area to force the slope of $\tau_\varphi^{-1}(T)$ to match that predicted by Eq.~(\ref{eq:AA}), giving $\rho_{xx}\sim 2800~\Omega$, mean free path $\sim 26$ nm, diffusion constant $D=131\pm 10~\rm{cm}^2/\rm{s}$, and the slope $A_s \approx 22~\rm{K}^{-1}\rm{ns}^{-1}$.  The black dashed line in Fig.~\ref{fig:2}(a) shows the plot of eq.~(\ref{eq:AA}) for the rescaled geometry, which closely matches the rescaled data and extrapolates to a finite value corresponding to the decoherence time $\sim 20$ ps as $T\rightarrow 0$. Data presented in the remainder of this paper is based on the rescaled area, but we point out that the alternative approach (ii) to treat the parameter $A$ does not alter the qualitative result of the following analysis of the quantum transport data.

Figure~\ref{fig:2}(b) shows $\tau_{\varphi}$, extracted from the curvature of rescaled magneto-conductance, for different values of $B_\parallel$ ($T=450~\rm{mK}$). The observed dependence $\tau_\varphi^{-1}(B_\parallel)$ features three characteristic regimes. For the intermediate fields (II), the polarisation of spin-scattering impurities at $g_i\mu_B B_\parallel \gtrsim k_B T$, suppresses spin-flip scattering and decreases $\tau_\varphi^{-1}$. This prolongation of phase coherence by the in-plane field is a smoking gun for spin-flip scattering in the system.  The suppression of $\tau_\varphi^{-1}$ is by a factor of approximately two before it begins to saturate at high field, indicating the entry into region III.
The high-field saturation may be explained by flexural deformation of graphene out of the plane, resulting in randomly-varying flux from the in-plane magnetic field \cite{FolkRipples}; by $g\sim 0$ magnetic moments that are not polarised by magnetic field; or by spin-orbit interaction.  This regime will be the subject of future work.

In the analysis that follows, we focus on the lowest-field regime (I), where the decoherence rate extracted using Eq. (\ref{eq:1}) shows a non-monotonic behaviour.   Early studies of WL in dilute magnetic alloys demonstrated a `spin-memory' effect \cite{Larkin,Falko1991} in the way the spinful scatterers reduce the size of the WL correction to conductivity.  In summary, the exchange interaction between a spinful impurity and the electron spin {\bf s}, via a perturbation $J\bf{s\cdot S} \delta(\bf{r-r}_i)$, affects the electrons scattering in two ways: (a) by scattering an electron without flipping its spin and (b) by scattering with a spin flip. Although the spin flip process (b) always leads to the decoherence of electrons, process (a) leads only to additional scattering phases for a fixed spin configuration of impurities (relative to the electron spin).  These phases would in general be the same for the two reversed sequences of visited scatterers, hence, they would not suppress the interference correction to conductivity. However, if the spin configuration of impurities is randomly changing in time, e.g., by Korringa relaxation, electron waves travelling along closed paths in opposite directions and therefore arriving at the same scatterer at different times would experience randomly different conditions for a spin-conserving scattering (`ergodic regime'), hence acquiring random relative phase shifts that suppress their interference contribution \cite{Falko1991}.

\begin{figure}[h!]

\begin{center}

\subfigure[]{
\includegraphics [scale=0.75] {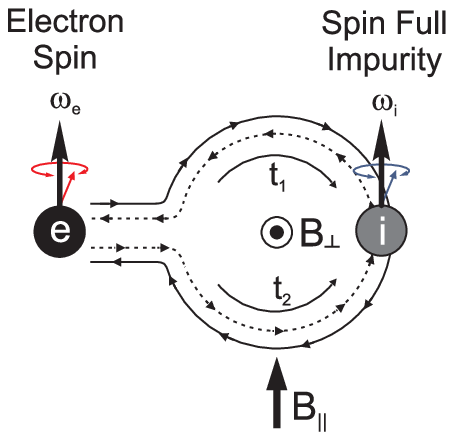}
\label{fig:subfig3a}
}
\subfigure[]{
\includegraphics [scale=0.65] {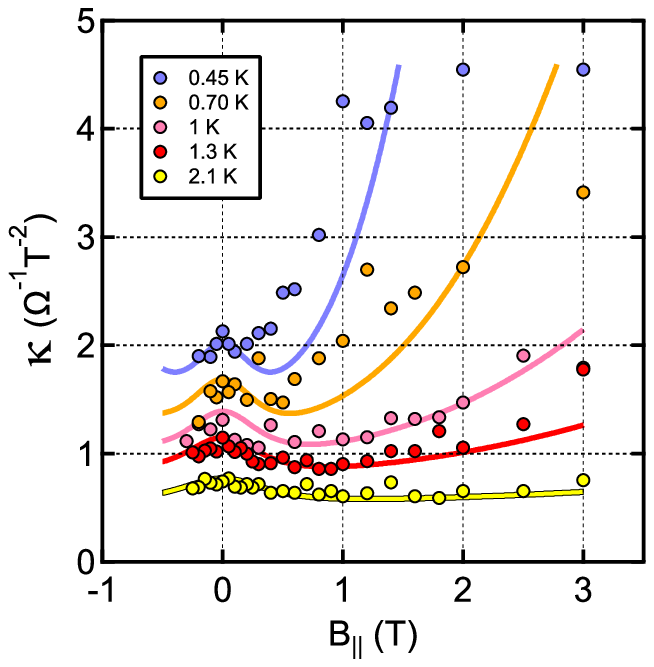}
\label{fig:subfig3b}
}

\end{center}
\caption{(a) Illustration of the influence of precession of the impurity spin on the weak localisation effect. (b) Comparison between the experimental values of the curvature of magnetoconductance at $T=0.45-2.1$~K with the theoretical values calculated using the parameters obtained using a fit at $T=2.1$~K.} \label{fig:3}
\end{figure}

In contrast to Korringa relaxation, spin precession of impurity spins does not necessarily randomise spin-dependent scattering conditions for diffusing electron waves. If electrons and scatterers have equal g-factors $g_i = g_e\approx 2$, their precessions in an external magnetic field are in phase and, because scattering conditions depend only on the relative orientation of the two spins, the `spin-memory' regime would also persist in a finite $B_\parallel$.
For scatterers with $g_i \ne g_e$, on the other hand, spin-dependent elastic scattering amplitudes for electrons following clockwise and anti-clockwise trajectories can be partly destroyed by even a small $B_\parallel$.  As sketched in Fig.~\ref{fig:3}(a), relative orientations of the spins of a scatterer and of the electron waves arriving along clockwise and anti-clockwise trajectories responsible for the interference correction to conductivity would typically deviate by an angle $(g_e-g_i)\mu_B  B_\parallel \tau_\varphi /\hbar$  and become randomised when $(g_e-g_i)\mu_B  B_\parallel > h\tau_\varphi^{-1}$.  This effect should be strongest in systems where the difference between the electron and scatterer $g$-factors is largest (in particular, in the systems where these two have opposite signs).  This leaves a range of magnetic field where precessional decoherence can fully develop (the ergodic regime) before the scatterers' spins become closely aligned with the external field ($B_\parallel >B_T$), thus diminishing the role of precessional dynamics.

In WL theory, the difference between these `spin-memory' and `ergodic' regimes is accounted for by numerical factors that appear in the relations between dephasing rates $\tau_{S,M}^{-1}$ of two-electron correlators, Cooperons, and the electron scattering rate from the spinful impurity, $\tau_{s}^{-1} = 2\pi\gamma n_i J^{2} S(S+1)$. Here, $\gamma=\frac{1}{2\hbar^{2}v}\sqrt{\frac{n_{e}}{\pi}}$ stands for the density of states of carriers and $v$ is Dirac velocity of electrons in graphene. The Cooperons are classified according to their total spin $S=0,1$ and its projection $M$ onto the direction of external magnetic field. These four Cooperons  $C_{S,M}$ are combined in the expression \cite{Larkin} for the WL correction to conductivity \cite{Note2},

\begin{equation}
\delta\sigma = \frac{e^2}{2\pi h}
\left[ C_{0,0} - C_{1,0} - C_{1,1} -C_{1,-1} \right]. \nonumber
\end{equation}
In the spin-memory regime,  $\tau_{0,0}^{-1}=2\tau_s^{-1} + \tau^{-1}_T$ and
$\tau_{1,M}^{-1}=\frac{2}{3}\tau_s^{-1} + \tau^{-1}_T$, whereas in the ergodic regime \cite{Falko1991} all Cooperons decay with the same rate, $\tau_{S,M}^{-1}=\tau_s^{-1} + \tau^{-1}_T$. Taking into account a possible difference between spin precession of the electrons and scatterers (which we do by changing the spin coordinates into the frame rotating with the frequency $\omega_e$ around the direction of external magnetic field $\bf{B}_\parallel$), we find using diagrammatic perturbation theory \cite{Kashuba} that the two Cooperons $C_{1,\pm 1}$ with $S=1$ and $M=\pm1$ are decoupled from all others and decay at the rate,
\begin{equation}
\tau^{-1}_{1,\pm1} = \frac{1}{\tau_{s}}\left[1 - \frac{\langle S_{z}^{2}\rangle \pm \langle S_{z}\rangle \bigl(1 - 2 n_{F}(\varepsilon_{\mp}) \bigr) }{S(S+1)} \right] +\tau_T^{-1},
\nonumber
\end{equation}
where $\varepsilon_{\mp}=\varepsilon\pm g_{i}\mu_{B}B/2$ takes into account that spin-flip scattering requires energy transfer to the scatterer.
At the same time, spin precession mixes the two Cooperons, $C_{0,0}$ and $C_{1,0}$. This mixing generates combined modes, which relax with the decoherence rates
\begin{equation}
\begin{split}
\tau^{-1}_{0,\pm} =
\left[1 + \frac{\langle S_{z}^{2} \rangle - \langle S_{z}\rangle [ n_{F}(\varepsilon_{+}) - n_{F}(\varepsilon_{-}) ] }{S(S+1)} \right] \tau_{s}^{-1}
\nonumber \\
\pm \sqrt{\tau^{-1}_{1,+1} \tau^{-1}_{1,-1}-(g_{e}-g_{i})^{2}\mu_{B}^{2}B_\parallel^{2}/\hbar^2} +\tau_T^{-1}.
\nonumber
\end{split}
\end{equation}
Note that, at $B=0$, $\langle S_{z}\rangle=0$ and $\langle S_{z}^{2} \rangle=\frac13 S(S+1)$, so that $\tau^{-1}_{1,\pm1}=\tau^{-1}_{0,-}=\frac23 \tau^{-1}_{s}+\tau_T^{-1}$ and $\tau^{-1}_{0,+}=2\tau^{-1}_{s}+\tau_T^{-1}$, corresponding to the relaxation rates of triplet and singlet Coperons in the spin-memory regime. At $B_\parallel >B_T$, $\tau^{-1}_{1,\pm1}=\tau_T^{-1}$, reflecting the restoration of phase coherence of diffusion of spin-polarised electrons.

Together, the four Cooperon modes \cite{Note2} yield:
\begin{equation}
\begin{split}
\delta\sigma = \frac{e^{2}}{2\pi h} \int d\varepsilon \sum_{\alpha=\pm}
n_{F}'(\varepsilon_{\alpha})
\left[
 \ln\frac{\tau_{1,\alpha}}{\tau_{iv}}
+ A_{\alpha}\ln\frac{\tau_{0,-}}{\tau_{0,+}}
\right],  \\ \nonumber
A_{\pm} = (\tau_{1,\pm1}^{-1}-\tau_T^{-1})/(\tau_{0,+}^{-1}-\tau_{0,-}^{-1}); \nonumber
\end{split}
\end{equation}
and magneto-conductance (measured as a function of $B_\perp$ for fixed $B_\parallel$),
\begin{equation}
\begin{split}
\sigma(B_\perp,B_\parallel)=\sigma(0,B_\parallel) -\frac{e^{2}}{2\pi h} \int d\varepsilon \sum_{\alpha=\pm}
n_{F}'(\varepsilon_{\alpha}) \\ \nonumber
\times \left(
  F(\frac{B_\perp}{B_{1,\alpha}})
+ \left[ F\left( \frac{B_\perp}{B_{0,-}} \right) - F\left(\frac{B_\perp}{B_{0,+}}\right) \right] A_{\alpha}
\right). \nonumber \\
F(z)=\ln z+\psi(\frac12+\frac{1}{z}); \quad B_{\beta,\alpha}=\frac{\hbar/4e}{D\tau_{\beta,\alpha}}, \nonumber
\end{split}
\end{equation}
where $n_{F}'(\varepsilon)\equiv\partial n_F(\varepsilon) \partial\varepsilon$
is a derivative of the Fermi distribution function, $\psi$ is the digamma function, $\tau^{-1}$ is the momentum relaxation rate that determines the classical Drude conductivity and the diffusion coefficient, $D=\frac12 v^2\tau$ (note that for monolayer graphene studied in this work, $D\approx$131 cm$^2$/s), and $\tau_{iv}$ is the intervalley scattering time \cite{McCannFalko,Note2}.

A non-monotonic dependence of decoherence rate on the in-plane magnetic field follows from the calculated magneto-conductance curvature,
\begin{multline}
\kappa= \frac{2e^{2}}{3\pi h}
\int d\varepsilon\,\sum_{\alpha=\pm}\bigl(-n_{F}'(\varepsilon_{\alpha})\bigr) \times
\\
\times \Biggl\{
\left(\frac{D\tau_{1,\alpha}}{\hbar/e}\right)^{2}
+
(\tau_{1,\pm1}^{-1}-\tau_T^{-1})(\tau_{0,+}+\tau_{0,-}) \frac{D\tau_{0,+}}{\hbar/e} \frac{D \tau_{0,-}}{\hbar/e}
\Biggr\}
\label{eq:ddsigmafin}
\end{multline}
Here, we used the expansion $F(z)\approx z^{2}/24 + O[z^{3}]$.

Eq.~(\ref{eq:ddsigmafin}) gives a semi-quantitative description of the observed magneto-conductance curvature that captures its qualitative features over a wide range of temperatures $0.45-2.1$~K and in-plane magnetic fields (0-3~T) [Fig.~3b]. Here, the following protocol was used for a single fit that yielded all five curves shown in Fig.~3b:

\begin{enumerate}
\item The temperature dependence of magneto-conductance curvature $\kappa$ at $B_\parallel = 0$ is used to extract $\tau_s^{-1} \approx 77$ ns$^{-1}$ and the parameter $A_t$ $\approx 24$ ns$^{-1}$K$^{-1}$ from the fit to our theory to be compared with $A_s \approx 22$ ns$^{-1}$K$^{-1}$ from eq. (\ref{eq:AA}). The red line in Fig. \ref{fig:2}a is the theoretical temperature dependence, which nearly coincides with the straight dashed line given by eq. (\ref{eq:AA}) and perfectly matches the experimental points.
\item  The dependence of $\kappa$ on $B_\parallel$ at high temperature, $T=2.1$ K [bottom dataset in Fig.~3(b)], is used to determine the average $g$-factor of the scatterers, which returned the value $g_i = -0.84\pm0.25$.
\end{enumerate}

The parameters determined above are sufficient to calculate the $B_\parallel$-dependence of the magneto-conductance curvature for lower temperatures, down to $T=$450 mK, which are then  compared with the experimentally measured values [Fig.~3(b)]. The theoretically calculated curvature (lines) captures the observed non-monotonic behaviour and a shift of the low-field anomaly towards smaller $B_\parallel$ at lower temperatures \cite{Note3}.

To conclude, we have demonstrated that the excess decoherence rate, observed earlier at low temperatures in epitaxial graphene sublimated on SiC \cite{Lara,Jobst}, is caused by the spin-flip scattering of electrons from spinful impurities. These spinful scatterers have an average $g$-factor $g_i=-0.84 \pm 0.25$, very different from the free-electron $g$-factor in graphene, $g_e\approx 2$, which enabled us to observe the influence of precession of the impurities' spins on the WL effect. The large (and negative) $g$-factor for the impurity spin implies a strong atomic spin-orbit coupling in the magnetic moment formation.   This may indicate that these spins are in the surface states underneath the graphene layer, possibly originating in Si substitutions of carbon atoms in the interfacial layer \cite{Hass}. The presence of such spinful scatterers on the SiC surface, directly accessible for graphene electrons, explains the short spin coherence length observed in spin-valve experiments on a similar material \cite{Maassen2013}.

\section*{Acknowledgements}
We thank I.~Aleiner, L.~Glazman, C. Marcus, B. van Wees, and A.~MacDonald for useful discussions. This work has been supported by NSERC, CFI, QMI, the EC Graphene Flagship (CNECT-ICT-604391), EMRP project GraphOhm and ERC Synergy Grant Hetero2D.

\end{document}


\title{Influence of impurity spin dynamics on quantum transport in epitaxial graphene\\ Supplemental Material}
\author{Samuel Lara-Avila}

\author{Oleksiy Kashuba}

\author{Joshua A. Folk}

\author{Silvia L\"{u}scher}

\author{Sergey Kubatkin}

\author{Rositza Yakimova}

\author{T.J.B.M. Janssen}

\author{Alexander Tzalenchuk}

\author{Vladimir Fal'ko}

\maketitle

\section{Theoretical model for the decoherence anomaly in 2D materials with magnetic defects in the in-plane magnetic field}

To explain the theoretical basis for the  anomaly in $\kappa(B_{\parallel})$ discussed in the main text of the Letter, we consider an electron wave propagating along a closed-loop trajectory, scattering from disorder,
\begin{equation}
V(\mathbf{r}) =  \sum_{\mathbf{r}_{U,j}} U \delta(\mathbf{r}-\mathbf{r}_{U,j}) + \sum_{\mathbf{r}_{J,j}}  J\,\mathbf{S}_{j}\cdot \mathbf{s}\,\delta(\mathbf{r}-\mathbf{r}_{J,j}),
\end{equation}
Zeeman splitting for electrons and impurities are taken into account by
$$ -g_{e}\mu_{B}B_{\parallel}\sigma; \quad -g_{i}\mu_{B}B_{\parallel}(\mathbf{S}_{j})_{z},$$
where $z$-projection $\sigma$ of electron's spin takes values $+^{1}\!/_{2}$($\uparrow$ or $+$) or $-^{1}\!/_{2}$ ($\downarrow$ or $-$); $g$-factor of the electron $g_{e}\approx-2.002$, and $\mu_{B}=|e|\hbar/2m c$ is Bohr magneton.

In the WL theory \cite{Larkin}, the interference correction, $\delta\sigma$, to conductivity is expressed as in term of two-electrons propagators, 'Cooperons' $\mathcal{C}_{\sigma\sigma';\eta\eta'}(\varepsilon,\mathbf{q})$, as
\begin{multline}
\delta\sigma = \frac{e^2}{2\pi h} \left[ C_{0,0} - C_{1,0} - C_{1,1} -C_{1,-1} \right]; \\
C_{S,M} = \zeta_{S,M;\sigma\sigma'} \hat{C} \zeta_{S,M;\sigma\sigma'}^\mathsf{T} \\
\hat{C}  = -\gamma v_{F}^{2}\tau^{3}\int d\varepsilon d^{2}\mathbf{q}
n_{F}'(\varepsilon_{\sigma'})
\hat{\mathcal{C}}(\varepsilon,\mathbf{q}).
\label{eq:WLconductivity}
\end{multline}
corresponding to the diagramatic perturbation theory represenation shown in Fig.~1. Here, Clebsch-Gordan coefficients $\zeta_{S,M;\sigma\sigma'}=\langle S,M|^{1}\!/_{2},\sigma;^{1}\!\!/_{2},\sigma'\rangle$ select from the Cooperon matrix $\hat{\mathcal{C}} \equiv (\mathcal{C}_{\sigma\sigma';\eta\eta'})$
the singlet ($S=0$) and triplet ($S=1$, $M=-1,0,1$) components defined in terms of the total spin carried by the two-electron propagator and its projection onto the external magnetic field $\mathbf{B}_\parallel$. Also, $\varepsilon_{\sigma}=\varepsilon - \sigma g_{i}\mu_{B}B_{\parallel}$ \cite{FX}, and $n_{F}'(\varepsilon)\equiv\partial n_F(\varepsilon) /\partial\varepsilon$ is a derivative of the Fermi distribution function.

To mention, in graphene, the weak localisation behaviour may be additionally affected by the Berry phase of electrons arising from the sublattice composition of their wave functions on the honeycomb lattice of carbons \cite{McCannFalko}. This would lead to the antilocalisation behavior in structures with a long intervalley scattering time, turned into WL behaviour for long trajectories with the time of diffusive flight longer than the intervalley scattering time. In the materials studies here, intervalley scattering time has been found to be shorter than the dephasing time.  This permits us to study only valley-singlet Cooperons and ignore a possible sublattice and valley structure of elastic (spin-less) disorder in graphene, hence, to derive the form of the Cooperon equations affected by spin-flip scattering using a simplified model for electrons with structure-less (on the unit cell scale) Bloch functions.

\begin{figure}
\centering
\includegraphics[scale=0.55]{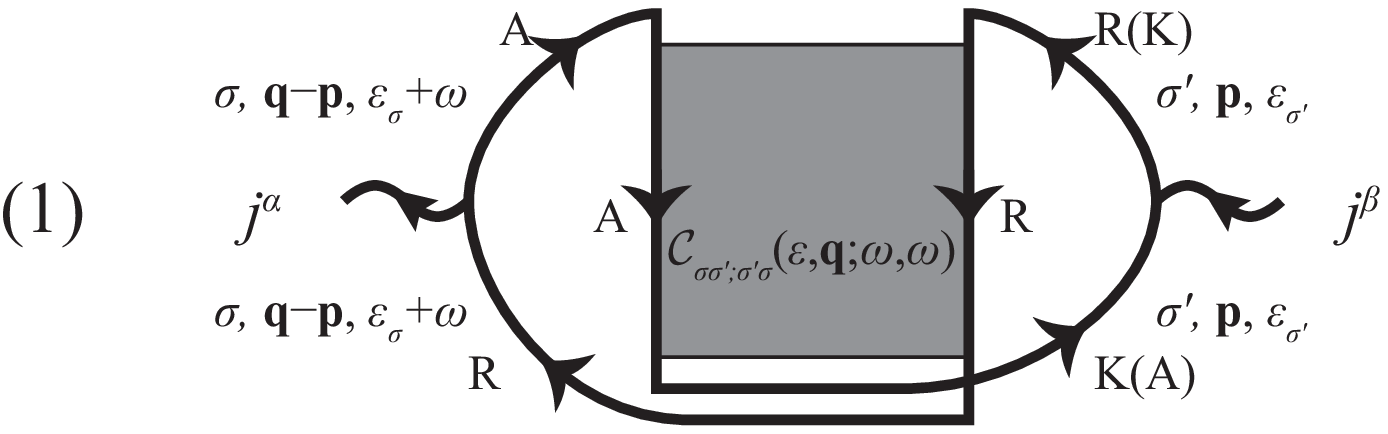}
\includegraphics[scale=0.55]{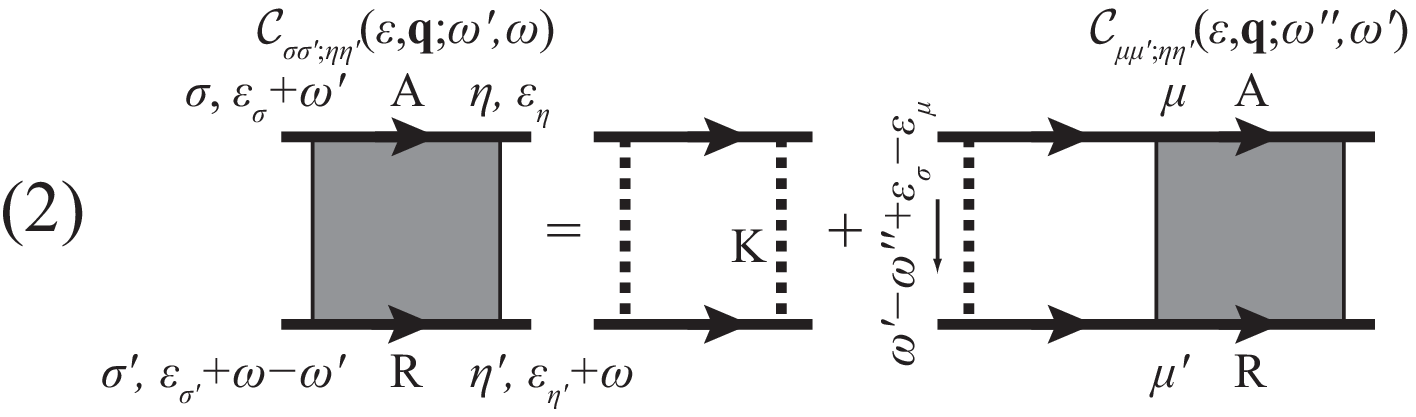}
\includegraphics[scale=0.55]{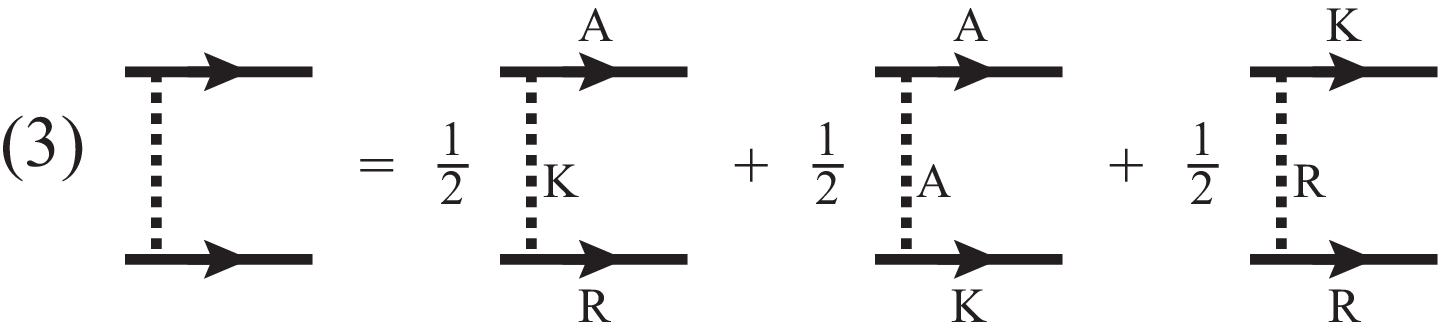}
\caption{Line (1): The diagram for Kubo susceptibility corresponding to the WL correction to the conductivity expressed in terms of the Cooperon $\mathcal{C}_{\sigma\sigma';\eta\eta'}(\varepsilon,\mathbf{q};\omega',\omega)$.
$j^{\alpha(\beta)}$ are current operators, the Keldysh components as well as spin, momentum and energy transferred through electron Green's function are denoted.
Line (2): Diagrammatic representation of Bethe-Salpeter equation for Cooperon $\mathcal{C}_{\sigma\sigma';\eta\eta'}(\varepsilon,\mathbf{q};\omega',\omega)$.
The parametrization of the Cooperon frequencies is shown together with spin states.
Line (3): Keldysh structure of the equation's kernel.}
\label{fig:kubobethesalp}
\end{figure}

\begin{figure}
\centering
\includegraphics[scale=0.55]{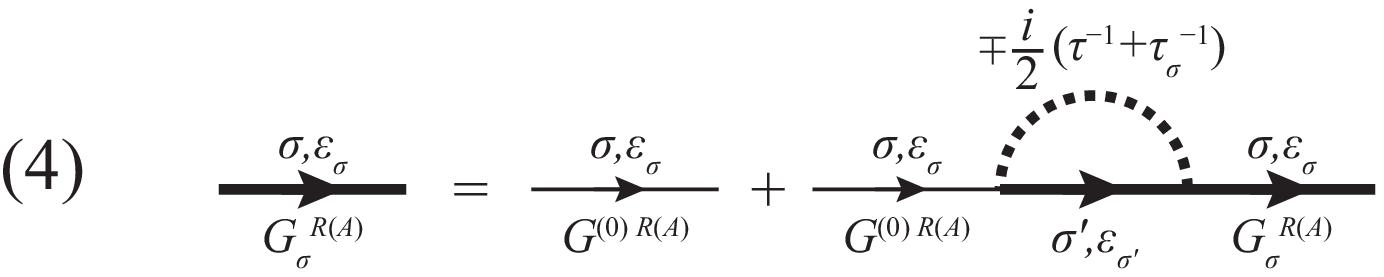}
\includegraphics[scale=0.55]{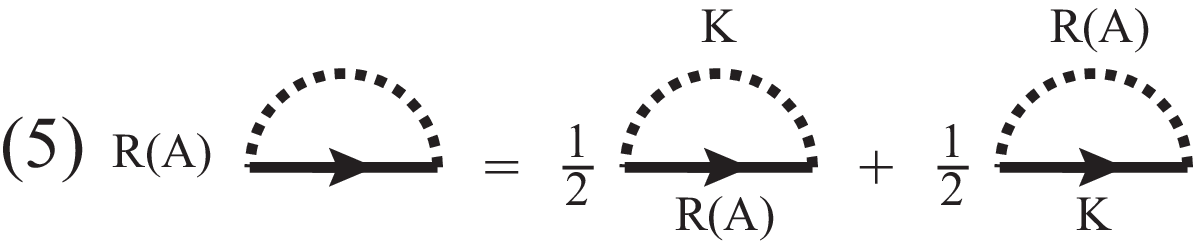}
\caption{Line (4): Diagrammatic representation of Dyson equation for single electron Green's function.
The last diagram contains the selfconsistent self-energy calculated in the main order of $1/vp_{F}\tau$ expansion.
Line (5): Keldysh structure of the self-energy.}
\label{fig:greenfunc}
\end{figure}

The diagrammatic form of the Bethe-Salpeter equation for the Cooperon matrix  $\mathcal{C}$ is shown in Fig.~1(2).
Thick solid lines in Fig.~1  denote disorder-averaged single-electron Green's functions $G_{\sigma}$ (obtained from the solution of Dyson equation in  Fig.~2),
\begin{equation}
\begin{split}
G^{R/A}_{\sigma} &= \bigl(\varepsilon_{\sigma} \!-\! v_{F}\xi_{\mathbf{p}} \!+\! g_{e}\mu_{B}B_{\parallel}\sigma
\pm \frac{i}{2}(\tau^{-1}\!+\!\tau_{\sigma}^{-1})\bigr)^{-1};
\\
G^{K}_{\sigma} &= \bigl(1-2n_{F}(\varepsilon_{\sigma})\bigr) \bigl(G^{R}_{\sigma}-G^{A}_{\sigma}\bigr); \\
\end{split}
\end{equation}
where
\begin{equation}
\begin{split}
\tau_{\sigma}^{-1} = \left[1 - 2\sigma M_{1} \bigl(1-2n_{F}(\varepsilon_{-\sigma})\bigr)\right]\tau_{s}^{-1};
\\
M_{n}=\langle S_{z}^{n}\rangle/S(S+1); \quad \xi_{\mathbf{p}}\approx v_{F}(|\mathbf{p}|-p_{F}),
\end{split}
\end{equation}
and $\sigma$ is electron's spin projection on the direction of in-plane magnetic field.
Here
$$\langle S_{z}^{n}\rangle = [Z(a)]^{-1}\partial^{n}_{a}Z(a),$$
where $a=g_{i}\mu_{B}B_{\parallel}/kT$, and
$$Z(a)=\sum_{S_{z}=-S}^{S}e^{aS_{z}}$$ is a statistical sum for the spin ensemble of paramagnetic scatterers.

The disorder correlation function is represented in Figs.~1 and 2 by the dashed lines. Here, we assume short-range spatial correlations and include the following elements in the description of disorder listed in Fig.~3: \\
(a) spatial correlator has \emph{only Keldysh component}
\begin{equation*}
(2\pi\gamma\tau)^{-1} \delta(\omega)\delta(\mathbf{r}-\mathbf{r}')
\end{equation*}
for the short range spinless disorder (the first term in $V(\mathbf{r})$), which is parametrised by the momentum relaxation rate, $\tau^{-1}=2\pi\gamma n_{U}U^{2}$ and the electron density of states $\gamma$;\\
(b) correlators of $z$-spin components for paramagnetic defects that characterise spin-dependent scattering of electron without spin flip,
\begin{equation*}
\frac{1}{2\pi\gamma\tau_s} \langle \mathrm{T}_{K} S_{z}(t)S_{z}(t')\rangle  \delta (\mathbf{r}-\mathbf{r}'),
\end{equation*}
$\tau_{s}^{-1}=2\pi\gamma n_{J}J^{2}$, which is independent on the positions $t$ and $t'$ on the Kendysh contour and therefore also has \emph{only Keldysh component}, Fourier transform of which is
\begin{equation*}
\frac{\langle S_{z}^{2}\rangle}{2\pi\gamma\tau_s}\delta(\omega) \delta (\mathbf{r}-\mathbf{r}');
\end{equation*}
(c,d) spin-flip correlators for paramagnetic defects, defined on the Keldysh time contour,
\begin{equation*}
\frac{1}{2\pi\gamma\tau_s}\langle \mathrm{T}_{K} S_{+}(t)S_{-}(t')\rangle  \delta (\mathbf{r}-\mathbf{r}'),
\end{equation*}
where correlator $D(t,t')=\langle \mathrm{T}_{K} S_{+}(t)S_{-}(t')\rangle$, written as
a matrix in Keldysh space \cite{LifshitzPitaevskii}, has the following components \cite{FX},
\begin{equation}
\begin{split}
D^{R/A}(\omega)&=2i\langle S_{z}\rangle (\omega-g_{i}\mu_{B}B\pm i0)^{-1},
\\
D^{K}(\omega)&=4\pi [S(S+1)-\langle S_{z}^{2}\rangle]\delta(\omega-g_{i}\mu_{B}B).
\end{split}
\end{equation}

\begin{figure}
\centering
\includegraphics[scale=0.55]{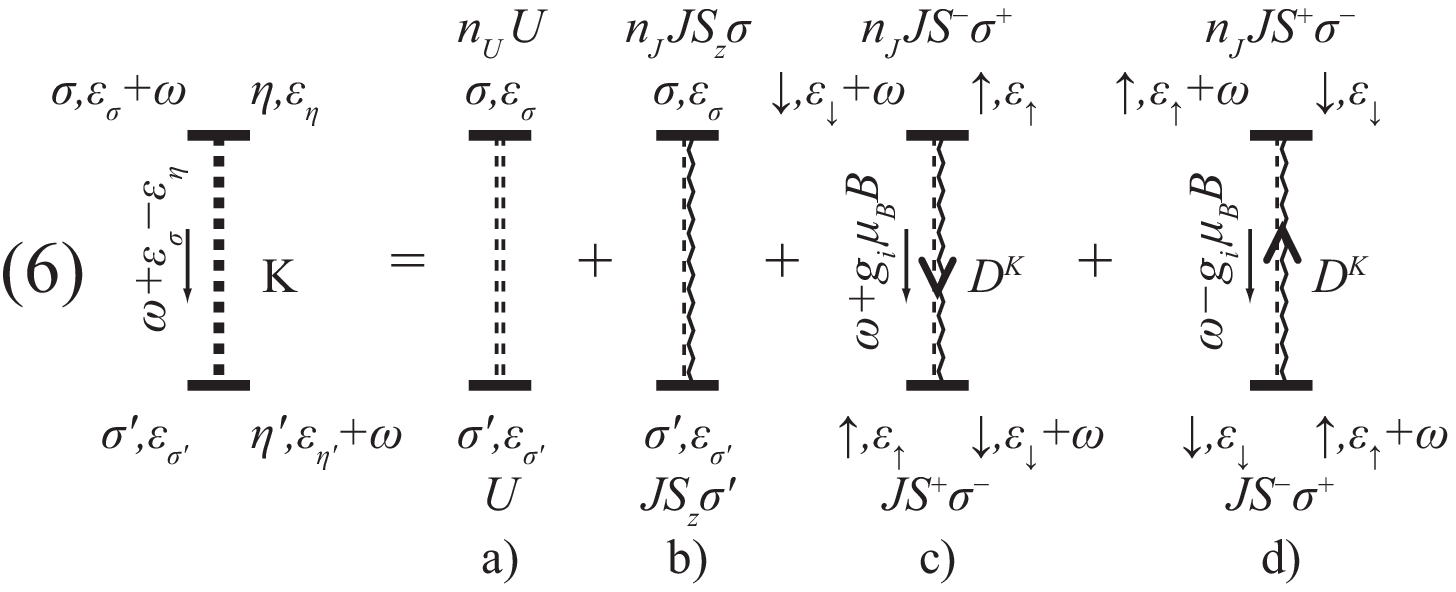}
\includegraphics[scale=0.55]{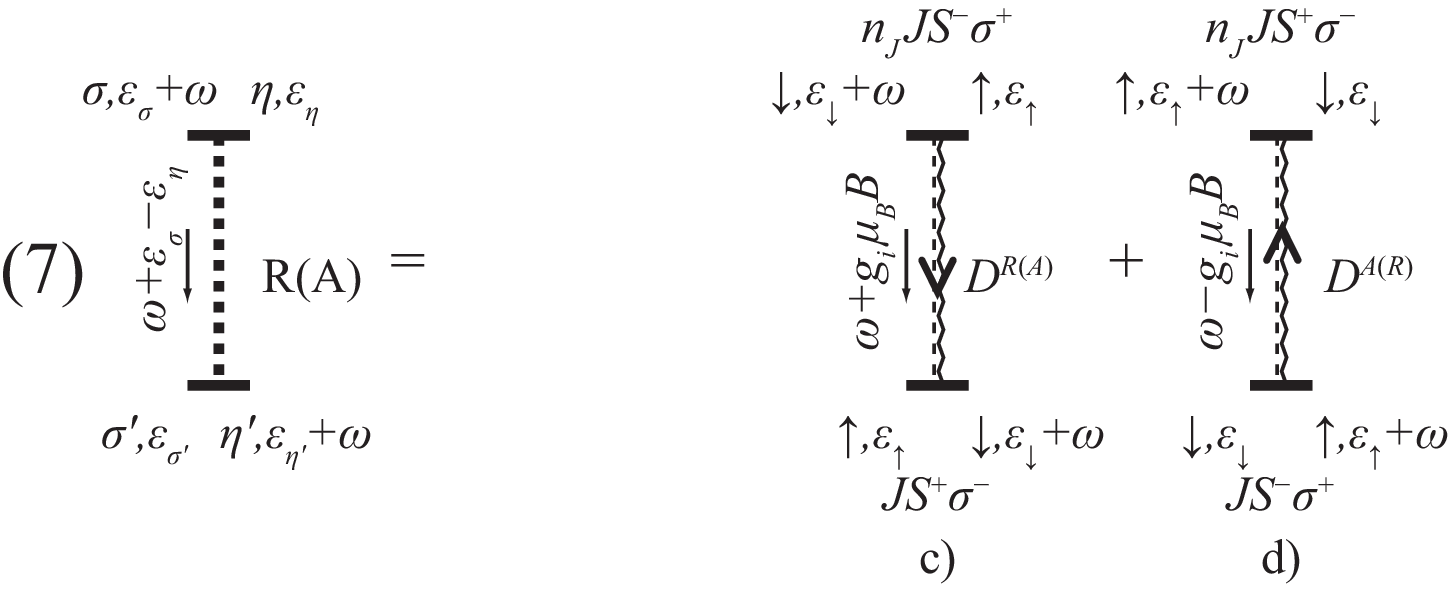}
\caption{Line (6): The Keldysh element of the diagram corresponding to the two-electrons scattering on the impurity (after averaging over disorder).: (a) dominating spin-independent scattering, (b) elastic spin-dependent without spin-flip, and (c,d) spin-flip scattering.
Line (7): The retarded (advanced) element of the diagram corresponding to the two-electrons spin-flip scattering on the impurity (c,d).
The parameters of incoming and outgoing Green's function are denoted.}
\label{fig:scattline}
\end{figure}

The implementation of Keldysh functions of the defects' spins in the perturbation theory diagrams is shown in Figs.~1 and 2.
Summing all three possible combinations of Keldysh components noted in Fig.~1, ($G^{A}D^{K}G^{R}$, $G^{A}D^{A}G^{K}$, and $G^{K}D^{R}G^{R}$), where the frequency argument of the spin corellator is $D^{R/K/A}\equiv D^{R/K/A}(\omega+g_{i}\mu_{B}B)$, we get the contribution to the kernel of Bethe-Salpeter equation for the Cooperon, including all  spin exchange $\downarrow\uparrow\to\uparrow\downarrow$ processes (Fig.~3 c) and d)):
\begin{multline*}
\frac{1}{4\pi\gamma\tau_{s}}\Bigl( D^{K} + [1-2n_{F}(\varepsilon_{-}+\omega)](D^{A}-D^{R}) \Bigr)\\
=\frac{1}{\gamma\tau_{s}}
\Bigl(
S(S+1) \!-\! \langle S_{z}^{2}\rangle \!-\! [1 \!-\! 2n_{F}(\varepsilon_{-})]\langle S_{z}\rangle
\Bigr)
\delta(\omega).
\end{multline*}
We note that in the approximation where only the Cooperons $\hat{\mathcal{C}}(\varepsilon,\mathbf{q};\omega',\omega)\propto\delta(\omega')$ are taken into account, the energy transferred through the impurity spin can be only $\pm g_{i}\mu_{B}B_{\parallel}$, where the sign depends on whether the impurity transfers spin $1$ or $-1$.

After this, the equation for the Cooperon takes the form
\begin{multline}
\left(
D\left(-i\hbar\partial_{\mathbf{r}}-\frac{2e}{c}\mathbf{A}_{\perp}\right)^{2}+\hat{R}
\right) \hat{\mathcal{C}}(\varepsilon,\mathbf{r})=\frac{1}{2\pi\gamma\tau^{2}}\delta(\mathbf{r});  \\
R_{\sigma\sigma';\eta\eta'} = \delta_{\sigma\eta}\delta_{\sigma'\eta'} \Bigl((\tau_{\sigma}^{-1}+\tau_{\sigma'}^{-1})/2 + \tau_{T}^{-1}\Bigr)
+
\\
+ \delta_{\sigma+\sigma',\eta+\eta'}\Bigl(
- 4\sigma\sigma' \tau_{s}^{-1} M_{2}
+ |\sigma-\eta|\tau_{\sigma}^{-1} +\\
+ i(\sigma-\sigma')(g_{e}-g_{i})\mu_{B}B \Bigr),
\label{eq:BSe}
\end{multline}
where $D=v_{F}^{2}\tau/2$ is a diffusion coefficient, $\hat{\mathcal{C}}(\varepsilon,\mathbf{r})$ is the Fourier transform of $\hat{\mathcal{C}}(\varepsilon,\mathbf{q})$ and $\mathbf{A}_{\perp} = (0,B_{\perp}x)$ is the vector potential of the perpendicular magnetic field, which is assumed to be small ($B_{\perp}\ll B_{\parallel}$). Note that here we have already used a simplification of the Cooperon equations in graphene by assuming that both decoherence time and spin flip time in the material are longer than the intervalley scattering time, $\tau_s, \tau_\varphi \gg \tau_{iv}$, so that these Cooperons have to be understood as valley-singlet Cooperons \cite{McCannFalko}  (one does not need to involve valley-triplet Cooperons in the analysis).

Diagonalization of the matrix $\hat{R}$ produces Cooperons $C_{1,\pm 1}$ decoupled from all others with decay rates
\begin{equation}
\tau^{-1}_{1,\pm1} = \left[1 - M_{2} \mp M_{1} \bigl(1 - 2 n_{F}(\varepsilon_{\mp}) \bigr) \right]\tau_{s}^{-1} + \tau_T^{-1},
\label{eq:r1pm}
\end{equation}
where $\varepsilon_{\mp}$ takes into account that spin-flip scattering requires energy transfer to the scatterer.
At the same time, spin precession as well as spin-flip processes mix up two Cooperons, $C_{0,0}$ and $C_{1,0}$.
This mixing generates combined modes, which relax with the decoherence rates
\begin{equation}
\begin{split}
\tau^{-1}_{0,\pm} =
\left[1 + M_{2} - M_{1} \bigl( n_{F}(\varepsilon_{+}) - n_{F}(\varepsilon_{-}) \bigr) \right] \tau_{s}^{-1}
\\
\pm \sqrt{\tau^{-1}_{1,+1} \tau^{-1}_{1,-1}-(g_{e}-g_{i})^{2}\mu_{B}^{2}B_\parallel^{2}/\hbar^2} +\tau_T^{-1}.
\label{eq:r0pm}
\end{split}
\end{equation}
Note that, at $B_{\parallel}=0$, $\langle S_{z}\rangle=0$ and $\langle S_{z}^{2} \rangle=\frac13 S(S+1)$, so that $\tau^{-1}_{1,\pm1}=\tau^{-1}_{0,-}=\frac23 \tau^{-1}_{s}+\tau_T^{-1}$ and $\tau^{-1}_{0,+}=2\tau^{-1}_{s}+\tau_T^{-1}$, corresponding to the relaxation rates of triplet and singlet Cooperons in the spin-memory regime. At $B_{\parallel}\gg B_{T}$, $\tau^{-1}_{1,\pm1}=\tau_T^{-1}$, reflecting the restoration of phase coherence of diffusion of spin-polarized electrons.

Together, the four Cooperon modes obtained from Eq.~\eqref{eq:BSe} and substituted in Eqs.~\eqref{eq:WLconductivity} yield the WL correction to the conductivity at $B_\perp=0$
\begin{equation}
\begin{split}
\delta\sigma \!=\! \frac{e^{2}}{2\pi h} \! \int d\varepsilon \! \sum_{\alpha=\pm} \!
n_{F}'(\varepsilon_{\alpha})
\!\left[
\ln\frac{\tau_{1,\alpha}}{\tau_{iv}}
+ A_{\alpha}\ln\frac{\tau_{0,-}}{\tau_{0,+}}
\right],  \\ 
1/A_{\pm} \!=\! (\tau_{0,+}^{-1} \!-\! \tau_{0,-}^{-1})\tau^{(0)}_{1,\pm1}, \quad
1/\tau^{(0)}_{1,\pm1} \!=\! \tau^{-1}_{1,\pm1} \!-\! \tau_{T}^{-1};
\end{split}
\end{equation}
where $\tau_{iv}$ is the intervalley scattering time.

The magnetoconductance (defined as conductance variation as a function of $B_\perp$ for fixed $B_\parallel$) takes the form
\begin{equation}
\begin{split}
\sigma(B_\perp,B_\parallel)-\sigma(0,B_\parallel)= -\frac{e^{2}}{2\pi h} \int d\varepsilon \sum_{\alpha=\pm}
n_{F}'(\varepsilon_{\alpha}) \quad\\ 
\times \left(
  F(\frac{B_\perp}{B_{1,\alpha}})
+ \left[ F\left( \frac{B_\perp}{B_{0,-}} \right) - F\left(\frac{B_\perp}{B_{0,+}}\right) \right] A_{\alpha}
\right).  \\
F(z)=\ln z+\psi\left(\frac12+\frac{1}{z}\right); \quad B_{\beta,\alpha}=\frac{\hbar/4e}{D\tau_{\beta,\alpha}}, 
\end{split}
\label{eq:cond}
\end{equation}
where $\psi$ is digamma function.

By using the expansion $F(z)\approx z^{2}/24 + O[z^{3}]$, we obtain the formula for the magnetoresistance curvature (which gives a direct measure for the decoherence rate),
\begin{equation}
\begin{split}
\kappa=& -\frac{2e^{2}}{3\pi h}
\int d\varepsilon\,\sum_{\alpha=\pm}n_{F}'(\varepsilon_{\alpha})
\\
&\times \Biggl\{
\left(\frac{D\tau_{1,\alpha}}{\hbar/e}\right)^{2}
+
\frac{\tau_{0,+}+\tau_{0,-}}{\tau_{1,\alpha}^{(0)}} \frac{D\tau_{0,+}}{\hbar/e} \frac{D \tau_{0,-}}{\hbar/e}
\Biggr\}
\end{split}
\label{eq:curv}
\end{equation}
This result was used in the main text to reproduce the experimentally observed non-monotonic dependence of MR curvature on the in-plane magentic field and to explain the anomaly in the 'decoherence rate' that take place when  $|g_{e}-g_{i}|\mu_{B}B_{\parallel}\sim \hbar \tau_{s}^{-1}$.